\documentclass{article}
\usepackage{spconf,amsmath,epsfig}
\usepackage{amsfonts}
\usepackage{amssymb}
\usepackage{graphicx}
\graphicspath{{eps/}} \DeclareGraphicsExtensions{.eps}
\usepackage[all]{xy}
\usepackage{calc}
\usepackage{epic,eepic}
\usepackage{xcolor}
\usepackage{comment}

\def\x{{\mathbf x}}

\newcommand{\topspace}[1][4.25mm]{\vbox{\hbox{\vspace{#1}}}}

\title{MINIMUM-DELAY DECODING OF TURBO CODES FOR UPPER-LAYER FEC}
\name{Ghassan M. Kraidy and Valentin Savin
}
\address{\vspace{-8mm}\\ CEA-LETI, 17 Rue des Martyrs, 38054 Grenoble, France\\
\{ghassan.kraidy,valentin.savin\}@cea.fr}
\begin{document}
%
\maketitle
\begin{abstract}
In this paper we investigate the decoding of parallel turbo codes
over the binary erasure channel suited for upper-layer error
correction. The proposed algorithm performs ``on-the-fly'' decoding,
{\em i.e.} it starts decoding as soon as the first symbols are received.
This algorithm compares with the iterative decoding of codes defined
on graphs, in that it propagates in the trellises of the turbo code
by removing transitions in the same way edges are removed in a
bipartite graph under message-passing decoding. Performance
comparison with LDPC codes for different coding rates is
shown.
\end{abstract}
%
\section{Introduction}
The binary erasure channel (BEC) introduced by Elias
\cite{elias1955cnc} is one of the simplest channel models: a symbol
is either erased with probability $p$, or exactly received with
probability $1-p$.  The capacity of such a channel with a uniform
source is given by:
\begin{equation*}
C = 1 - p
\end{equation*}
Codes that achieve this capacity are called {\em Maximum-Distance
Separable} (MDS) codes, and they can recover the $K$ information
symbols from any $K$ of the $N$ codeword symbols. An MDS code that
is widely used over the BEC is the non-binary Reed-Solomon (RS)
code, but its block length is limited by the Galois field
cardinality that dramatically increases the decoding complexity. For
large block lengths, low-density parity-check (LDPC) codes
\cite{di2002fla} \cite{richardson_finite-length_2002}
\cite{richardson_design_2001} \cite{luby2001eec} and
repeat-accumu\-late (RA) \cite{pfister2005cae} codes with
message-passing decoding proved to perform very close to the channel
capacity with reasonable complexity. Moreover, ``rateless'' codes
\cite{luby2002lc} \cite{shokrollahi2006rc} that are capable of
generating an infinite sequence of parity symbols were proposed for
the BEC.  Their main strength is their high performance together
with linear time encoding and decoding. However, convolutional-based
codes, that are widely used for Gaussian channels, are less
investigated for the BEC.  Among the few papers that treat
convolutional and turbo codes \cite{berrou_near_1996} in this
context are \cite{tepe_turbo_1998} \cite{kurkoski_exact_2003}
\cite{kurkoski_analysis_2004} \cite{rosnes_turbo_2007}
\cite{jeong_w_lee_performance_2007}.

In practical systems, data packets received at the upper layers
encounter erasures.  In the Internet for instance, it is frequent to
have datagrams that are discarded by the physical layer cyclic
redundancy check (CRC) or forward error correction (FEC), or even by
the transport level user datagram protocol (UDP) checksums. Another
example would be the transmission links that exhibit deep fading of
the signal (fades of 10dB or more) for short periods. This is the
case of the satellite channel where weather conditions (especially
rain) severely degrades the channel quality, or even the mobile
transmissions due to terrain effect.  In such situations, the
physical layer FEC fails and we can either ask for re-transmission
(only if a return channel exists, and penalizing in broadcast/
multicast scenarios)
or use upper layer (UL) FEC.

In this paper, we propose a minimum-delay decoding algorithm for
turbo codes suited for UL-FEC, in the sense that the decoding starts
since the reception of the first symbols where a symbol could be a
bit or a packet. The paper is organized as follows: Section
\ref{sys_model} gives the system model and a brief recall of the
existing decoding algorithms. Section \ref{fly} explains the
minimum-delay decoding algorithm. Simulation results and comparisons
with LDPC codes are shown in Section \ref{results}, and Section
\ref{conclusion} gives the concluding remarks.

\section{System model and notations}
\label{sys_model} We consider the transmission of a
parallel turbo code \cite{berrou_near_1996} with rate $R_c=K/N$ over
the BEC.
An information bit sequence of length $K$ is fed to a recursive
systematic convolutional (RSC) code with rate $\rho=k/n$ to generate
a first parity bit sequence. The same information sequence is
scrambled via an interleaver $\Pi$ to generate a second parity
sequence. With half-rate RSC constituents, the resulting turbo code
has rate $1/3$. In order to raise the rate of the turbo code, parity
bits are punctured.  In this paper, we consider rate-$1/3$,
punctured rate-$1/2$ and punctured rate-$2/3$ turbo codes. The
decoding of turbo codes is performed iteratively using probabilities
on information bits, which requires the reception of the entire
codeword before the decoding process starts. For instance, the
soft-input soft-output (SISO) ``Forward-Backward'' (FB) algorithm
\cite{bcjr}, optimal in terms of {\em a posteriori} probability
(APP) on symbols, consists of one forward recursion and one backward
recursion over the trellis of the two constituent codes. As turbo
codes are classically used over Gaussian channels, a SISO algorithm
(the FB or other sub-optimal decoding algorithms) are required to
attain low error rates. Exchanging hard information between the
constituent codes using an algorithm such as the well-known Viterbi
Algorithm (VA) \cite{viterbi_error_1967} (that is a
Maximum-Likelihood Sequence Estimator (MLSE) for convolutional
codes) is harshly penalizing. However, in the case of the BEC, a
SISO decoding algorithm is not necessary. In fact, it has been shown
in \cite{kurkoski_analysis_2004} that the VA is optimal in terms of
symbol (or bit) probability on the BEC, which means that one can
achieve optimal decoding of turbo codes on the BEC without using
soft information. In other words, if a bit is known to (or correctly
decoded by) one trellis, its value cannot be modified by the other
trellis. Motivated by this key property, we propose a decoding
algorithm for turbo codes based on hard information exchange.

\section{On-the-fly decoding of turbo codes}
\label{fly}
The turbo code has two trellises that have $K$ steps
each, and one codeword represents a path in the trellises. In a goal
to minimize the decoding delay, we propose an algorithm that starts
decoding directly after the reception of the first bits of the
transmitted codeword. First, at every step of the trellises, if one
of the $n$ bits of the binary labeling is received ({\em i.e.} is
known), we remove the transitions that do not cover this bit.  If,
at some step in the trellis, all the transitions leaving a state
$e_i$ on the left are removed, we then know that no transition
arrives to this state at the previous step. Consequently, all the incoming
transitions to state $e_i$ from the left are removed.
Similarly, if - at some step - there are no transitions arriving to
a state $e_j$ on the right, this means that we cannot leave state
$e_j$ at the following step, and all the transitions outgoing from state
$e_j$ are removed. This way the information propagates in the
trellis and some bits can be determined without being received. This
algorithm is inspired by the message-passing decoding of LDPC codes
over the BEC, where transitions connected to a
variable node are removed if this variable is received.

Now at some stage of the decoding process, if an information bit is
determined in one trellis without being received, we set its
interleaved (or de-interleaved) counterpart as known and the same
propagation is triggered in the other trellis.  The information
exchange between the two trellises continues until propagation stops
in both trellises. This way we can recover the whole transmitted
information bits without receiving the whole transmitted codeword.

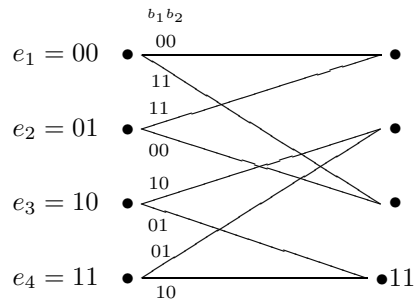
\begin{figure}[h!]
$$\xymatrix@C=30mm@R=5mm{
   e_1 = 00 \hspace{3mm} \bullet \ar@{-}'[]+R[r]+L^<<<<{00}^<<<<{\stackrel{b_1b_2}{\topspace[3mm]}}   \ar@{-}'[]+R[rdd]+L_<<<<{11\hspace{-1.7mm}} & \bullet \\
   e_2 = 01 \hspace{3mm} \bullet \ar@{-}'[]+R[ru]+L^<<<<{11}  \ar@{-}'[]+R[rd]+L_<<<<{00}  & \bullet \\
   e_3 = 10 \hspace{3mm} \bullet \ar@{-}'[]+R[ru]+L^<<<<{10}  \ar@{-}'[]+R[rd]+L_<<<<{01}  & \bullet \\
   e_4 = 11 \hspace{3mm} \bullet \ar@{-}'[]+R[ruu]+L^<<<<{01\hspace{-1.7mm}} \ar@{-}'[]+R[r]+L_<<<<{10}   & \bullet  11
}$$
 \caption{Transitions of the half-rate four-state RSC$(7,5)_8$ code.}
 \label{transitions}
\end{figure}

In the sequel, for the sake of clearness, we will only consider
parallel turbo codes built from the concatenation of two RSC codes
with generator polynomials $(7,5)$ in octal (the polynomial $\left(7\right)_8$
 being the feedback polynomial), constraint length
$L=3$, and coding rate $\rho=k/n=1/2$, code that has a simple
trellis structure with four states. The algorithm can be applied to
any parallel turbo code built from other RSC constituents. The
transitions of the RSC $(7,5)_8$ code between two trellis steps are
shown in Fig. \ref{transitions}.
As the code is systematic, the bit $b_1$ represents the information
bit, and the bit $b_2$ the parity bit. There are $2^k=2$ transitions
leaving and $2$ transitions arriving to each state.  The transitions
between two steps of the trellis can be represented by a $2^{L-1} \times 2^{ L-1}$ matrix ($4 \times
4$ matrix in this case). For the $(7,5)_8$ code for instance, the
transition table is given by:
\begin{center}
\begin{tabular}{|c|c|c|c|c|}
\cline{2-5}  \multicolumn{1}{c|}{}     & $e_1$ & $e_2$ & $e_3$ & $e_4$ \\
\hline $e_1$ &  00    &  X     &  11     &  X     \\
\hline $e_2$ &  11    &  X     &  00     &  X     \\
\hline $e_3$ &  X     &  10    &  X      &  01     \\
\hline $e_4$ &  X     &  01    &  X      &  10     \\ \hline
\end{tabular}
\end{center}
\vspace{2mm} where an $X$ means that the transition does not exist.
For the need of the proposed algorithm, we will use the transition
table of the code to build binary transition matrices $T_{\x\x}$,
$T_{b_1\x}$, and $T_{\x b_2}$ with $b_1,b_2 \in \{ 0,1
\}$ that contain the allowed transitions depending on the known
bits. These matrices will be stored at the decoder and used as {\em
look-up} tables throughout the decoding process. For instance, if
the two bits of the transition are unknown, we define the matrix:
\begin{equation*}
T_{\x\x}= \left[\begin{array}{cccc}
1 & 0 & 1 & 0\\
1 & 0 & 1 & 0\\
0 & 1 & 0 & 1\\
0 & 1 & 0 & 1
\end{array}\right]
\end{equation*}
where a one in position $(i,j)$ means that there is a transition
between state $e_i$ and state $e_j$, and a zero means that no
transition exists. However, if $b_1=0$ and $b_2$ is unknown, or if
$b_1$ is unknown and $b_2=0$, we define the following matrices corresponding to
the allowed transitions:
\begin{equation*}
T_{0\x}= \left[\begin{array}{cccc}
1 & 0 & 0 & 0\\
0 & 0 & 1 & 0\\
0 & 0 & 0 & 1\\
0 & 1 & 0 & 0
\end{array}\right]\!,\ \ T_{\x0}= \left[\begin{array}{cccc}
1 & 0 & 0 & 0\\
0 & 0 & 1 & 0\\
0 & 1 & 0 & 0\\
0 & 0 & 0 & 1
\end{array}\right]
\end{equation*}
We build the other matrices similarly. Note that there are a total
of $3^n$ matrices, each of size $2^{L-1} \times 2^{
L-1}$.
\pagebreak

\smallskip
{\em On-the-fly decoding algorithm}

\smallskip\noindent{\bf 1) Initialization step}.
We consider matrices $M_1(i)$ and $M_2(j)$ corresponding to
transitions at steps $i$ and $j$ of the two trellises of the
constituent codes. These matrices are initialized as follows: \small
$$\begin{array}{rr} M_{1,2}(0)= \left[\begin{array}{*{4}{c@{\,\,\,}}}
1 & 0 & 1 & 0\\
0 & 0 & 0 & 0\\
0 & 0 & 0 & 0\\
0 & 0 & 0 & 0
\end{array}\right]\!, & M_{1,2}(1)=\left[\begin{array}{*{4}{c@{\,\,\,}}}
1 & 0 & 1 & 0\\
0 & 0 & 0 & 0\\
0 & 1 & 0 & 1\\
0 & 0 & 0 & 0
\end{array}\right] \\
 & \\
M_{1,2}(K)= \left[\begin{array}{*{4}{c@{\,\,\,}}}
1 & 0 & 0 & 0\\
1 & 0 & 0 & 0\\
0 & 1 & 0 & 0\\
0 & 1 & 0 & 0
\end{array}\right]\!, & M_{1,2}(K+1)= \left[\begin{array}{*{4}{c@{\,\,\,}}}
1 & 0 & 0 & 0\\
1 & 0 & 0 & 0\\
0 & 0 & 0 & 0\\
0 & 0 & 0 & 0
\end{array}\right] \\
 & \\
\multicolumn{2}{c}{M_1(t)=M_2(t)= T_{\x\x},~~~~t=2, \dots, K-1}
\end{array}$$

\normalsize

The matrices at steps $0$ and $1$ (namely $M_1(0)$, $M_2(0)$,
$M_1(1)$ and, $M_2(1)$) represent the fact that any codeword starts
in the zero state. The matrices at steps $K$ and $K+1$ represent the
two steps required for trellis termination ({\em i.e.} ending in the
zero state).

\smallskip\noindent{\bf 2) Reception step}. Each time a bit $r \in \{0,1\}$ is received:
\begin{itemize}
\item If $r$ is an information bit, it is placed in appropriate positions in both trellises as $r_{2i}=r_{2j}=r$, where $j=\Pi \left( i \right)$.
We then compute:
\begin{equation*}
M_1(i)= M_1(i) \wedge T_{r\x}~~\mbox{and}~~M_2(j)= M_2(j) \wedge
T_{r\x}
\end{equation*}
where the $\wedge$ operator is a logical AND between corresponding
entries of the two matrices.  In other words, we only keep the
transitions in $M_1(i)$ with $b_1=r$.
\item If $r$ is a parity bit, we set $r_{2i+1}=r$ if $r$ belongs to
the first trellis, or $r_{2j+1}=r$ if $r$ belongs to the second one.
We then compute:
\begin{equation*}
M_1(i)= M_1(i) \wedge T_{\x r}~~\mbox{or}~~M_2(j)= M_2(j) \wedge
T_{\x r}
\end{equation*}
\end{itemize}

\noindent{\bf 3) Propagation step}. If either $M_1(i)$ or $M_2(j)$
has at least one all-zero column or one all-zero row, the algorithm
is able to propagate in either direction in either trellis using the
following rule:
\begin{itemize}
\itemsep 2mm
\item Let $d\in\{1,2\}$ represent the trellis indices and initialize a counter $t\in\{i,j\}$ representing the step index through each trellis.
\item Left propagation: an all-zero row with index $u$ in $M_d(t)$
generates an all-zero column with index $u$ in $M_d(t-1)$.
\item Right propagation: an all-zero column with index $v$ in $M_d(t)$
generates an all-zero row with index $v$ in $M_d(t+1)$.
\end{itemize}
If we get new all-zero columns or new all-zero rows at steps
$t\pm1$, we set $t\leftarrow t\pm1$ and continue the propagation
(Step 3).

\smallskip\noindent{\bf 4) Duplication step}. If during the propagation we get some $M_d(t) \subseteq T_{bx}$ ({\em i.e.} the value of the information bit
of the $t^{\mbox{\scriptsize th}}$ transition in the
$d^{\mbox{\scriptsize th}}$ trellis is equal to $b$), we proceed as
follows:
\begin{itemize}
\itemsep 2mm
\item If $M_1(t) \subseteq T_{b\x}$, we compute:
\begin{equation*}
M_2 \left( \Pi(t) \right) = M_2 \left( \Pi(t) \right) \wedge T_{b\x}
\end{equation*}
and then we propagate from $\Pi(t)$ in the second trellis (Step 3).

\item If $M_2(t) \subseteq T_{b\x}$, we compute:
\begin{equation*}
M_1 \left( \Pi^{-1}(t) \right) = M_1 \left( \Pi^{-1}(t) \right)
\wedge T_{b\x}
\end{equation*}
and then we propagate from $\Pi^{-1}(t)$ in the first trellis (Step
3).
\end{itemize}

\smallskip\noindent{\bf 5) New reception step}. If the propagation in both trellises stops, we go
back to step 2.

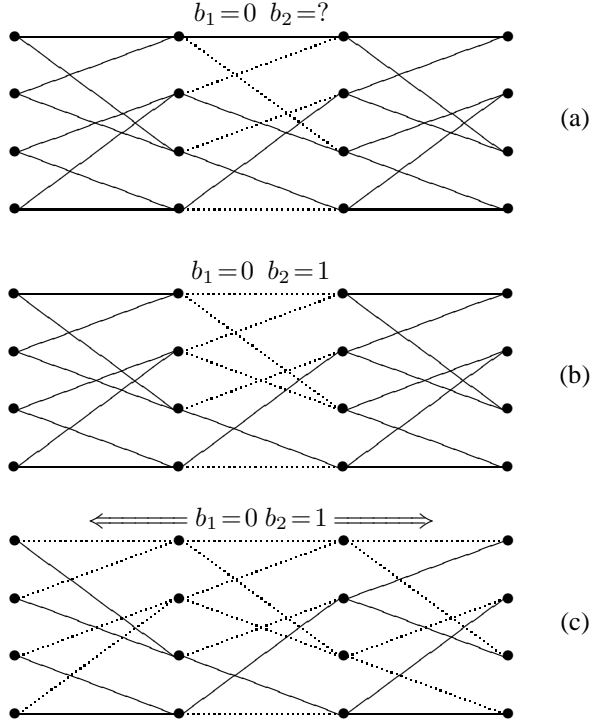
\begin{figure}[!th]
\vspace{-4mm}
$$\xymatrix@C=18mm@R=4mm{
   \bullet \ar@{-}'[]+<0.5mm,0mm>[r]+<-0.5mm,0mm>   \ar@{-}'[]+<0.5mm,0mm>[rdd]+<-0.5mm,0mm> & \bullet \ar@{-}'[]+<0.5mm,0mm>[r]+<-0.5mm,0mm>^{\raisebox{1mm}{$b_1\!=\!0~\ b_2\!= ?$}}   \ar@{.}'[]+<0.5mm,0mm>[rdd]+<-0.5mm,0mm> & \bullet \ar@{-}'[]+<0.5mm,0mm>[r]+<-0.5mm,0mm>   \ar@{-}'[]+<0.5mm,0mm>[rdd]+<-0.5mm,0mm> & \bullet\\
   \bullet \ar@{-}'[]+<0.5mm,0mm>[ru]+<-0.5mm,0mm>  \ar@{-}'[]+<0.5mm,0mm>[rd]+<-0.5mm,0mm>  & \bullet \ar@{.}'[]+<0.5mm,0mm>[ru]+<-0.5mm,0mm>  \ar@{-}'[]+<0.5mm,0mm>[rd]+<-0.5mm,0mm>  & \bullet \ar@{-}'[]+<0.5mm,0mm>[ru]+<-0.5mm,0mm>  \ar@{-}'[]+<0.5mm,0mm>[rd]+<-0.5mm,0mm>  & \bullet \\
   \bullet \ar@{-}'[]+<0.5mm,0mm>[ru]+<-0.5mm,0mm>  \ar@{-}'[]+<0.5mm,0mm>[rd]+<-0.5mm,0mm>  & \bullet \ar@{.}'[]+<0.5mm,0mm>[ru]+<-0.5mm,0mm>  \ar@{-}'[]+<0.5mm,0mm>[rd]+<-0.5mm,0mm>  & \bullet \ar@{-}'[]+<0.5mm,0mm>[ru]+<-0.5mm,0mm>  \ar@{-}'[]+<0.5mm,0mm>[rd]+<-0.5mm,0mm>  & \bullet\\
   \bullet \ar@{-}'[]+<0.5mm,0mm>[ruu]+<-0.5mm,0mm> \ar@{-}'[]+<0.5mm,0mm>[r]+<-0.5mm,0mm>   & \bullet \ar@{-}'[]+<0.5mm,0mm>[ruu]+<-0.5mm,0mm> \ar@{.}'[]+<0.5mm,0mm>[r]+<-0.5mm,0mm>   &  \bullet \ar@{-}'[]+<0.5mm,0mm>[ruu]+<-0.5mm,0mm> \ar@{-}'[]+<0.5mm,0mm>[r]+<-0.5mm,0mm>   & \bullet
} \hspace{5mm}\raisebox{-11mm}{(a)} $$
$$\xymatrix@C=18mm@R=4mm{
   \bullet \ar@{-}'[]+<0.5mm,0mm>[r]+<-0.5mm,0mm>   \ar@{-}'[]+<0.5mm,0mm>[rdd]+<-0.5mm,0mm> & \bullet \ar@{.}'[]+<0.5mm,0mm>[r]+<-0.5mm,0mm>^{\raisebox{1mm}{$b_1\!=\!0~\ b_2\!=\!1$}}   \ar@{.}'[]+<0.5mm,0mm>[rdd]+<-0.5mm,0mm> & \bullet \ar@{-}'[]+<0.5mm,0mm>[r]+<-0.5mm,0mm>   \ar@{-}'[]+<0.5mm,0mm>[rdd]+<-0.5mm,0mm> & \bullet\\
   \bullet \ar@{-}'[]+<0.5mm,0mm>[ru]+<-0.5mm,0mm>  \ar@{-}'[]+<0.5mm,0mm>[rd]+<-0.5mm,0mm>  & \bullet \ar@{.}'[]+<0.5mm,0mm>[ru]+<-0.5mm,0mm>  \ar@{.}'[]+<0.5mm,0mm>[rd]+<-0.5mm,0mm>  & \bullet \ar@{-}'[]+<0.5mm,0mm>[ru]+<-0.5mm,0mm>  \ar@{-}'[]+<0.5mm,0mm>[rd]+<-0.5mm,0mm>  & \bullet \\
   \bullet \ar@{-}'[]+<0.5mm,0mm>[ru]+<-0.5mm,0mm>  \ar@{-}'[]+<0.5mm,0mm>[rd]+<-0.5mm,0mm>  & \bullet \ar@{.}'[]+<0.5mm,0mm>[ru]+<-0.5mm,0mm>  \ar@{-}'[]+<0.5mm,0mm>[rd]+<-0.5mm,0mm>  & \bullet \ar@{-}'[]+<0.5mm,0mm>[ru]+<-0.5mm,0mm>  \ar@{-}'[]+<0.5mm,0mm>[rd]+<-0.5mm,0mm>  & \bullet\\
   \bullet \ar@{-}'[]+<0.5mm,0mm>[ruu]+<-0.5mm,0mm> \ar@{-}'[]+<0.5mm,0mm>[r]+<-0.5mm,0mm>   & \bullet \ar@{-}'[]+<0.5mm,0mm>[ruu]+<-0.5mm,0mm> \ar@{.}'[]+<0.5mm,0mm>[r]+<-0.5mm,0mm>   &  \bullet \ar@{-}'[]+<0.5mm,0mm>[ruu]+<-0.5mm,0mm> \ar@{-}'[]+<0.5mm,0mm>[r]+<-0.5mm,0mm>   & \bullet
} \hspace{5mm}\raisebox{-11mm}{(b)}$$
$$\xymatrix@C=18mm@R=4mm{
   \bullet \ar@{.}'[]+<0.5mm,0mm>[r]+<-0.5mm,0mm>^(.4){\stackrel{\,}{\topspace[2.5mm]}}="left"   \ar@{-}'[]+<0.5mm,0mm>[rdd]+<-0.5mm,0mm> & \bullet \ar@{.}'[]+<0.5mm,0mm>[r]+<-0.5mm,0mm>^{\raisebox{1mm}{$b_1\!=\!0\ b_2\!=\!1$}}="mid"   \ar@{.}'[]+<0.5mm,0mm>[rdd]+<-0.5mm,0mm> & \bullet \ar@{.}'[]+<0.5mm,0mm>[r]+<-0.5mm,0mm>^(.6){\stackrel{\,}{\topspace[2.5mm]}}="right"   \ar@{.}'[]+<0.5mm,0mm>[rdd]+<-0.5mm,0mm> & \bullet \ar@{=>}"mid";"left"_(.65){\mbox{\scriptsize }} \ar@{=>}"mid";"right"^(.65){\mbox{\scriptsize }}\\
   \bullet \ar@{.}'[]+<0.5mm,0mm>[ru]+<-0.5mm,0mm>  \ar@{-}'[]+<0.5mm,0mm>[rd]+<-0.5mm,0mm>  & \bullet \ar@{.}'[]+<0.5mm,0mm>[ru]+<-0.5mm,0mm>  \ar@{.}'[]+<0.5mm,0mm>[rd]+<-0.5mm,0mm>  & \bullet \ar@{-}'[]+<0.5mm,0mm>[ru]+<-0.5mm,0mm>  \ar@{-}'[]+<0.5mm,0mm>[rd]+<-0.5mm,0mm>  & \bullet \\
   \bullet \ar@{.}'[]+<0.5mm,0mm>[ru]+<-0.5mm,0mm>  \ar@{-}'[]+<0.5mm,0mm>[rd]+<-0.5mm,0mm>  & \bullet \ar@{.}'[]+<0.5mm,0mm>[ru]+<-0.5mm,0mm>  \ar@{-}'[]+<0.5mm,0mm>[rd]+<-0.5mm,0mm>  & \bullet \ar@{.}'[]+<0.5mm,0mm>[ru]+<-0.5mm,0mm>  \ar@{.}'[]+<0.5mm,0mm>[rd]+<-0.5mm,0mm>  & \bullet\\
   \bullet \ar@{.}'[]+<0.5mm,0mm>[ruu]+<-0.5mm,0mm> \ar@{-}'[]+<0.5mm,0mm>[r]+<-0.5mm,0mm>   & \bullet \ar@{-}'[]+<0.5mm,0mm>[ruu]+<-0.5mm,0mm> \ar@{.}'[]+<0.5mm,0mm>[r]+<-0.5mm,0mm>   &  \bullet \ar@{-}'[]+<0.5mm,0mm>[ruu]+<-0.5mm,0mm> \ar@{-}'[]+<0.5mm,0mm>[r]+<-0.5mm,0mm>   & \bullet
} \hspace{5mm}\raisebox{-11mm}{(c)}$$ 
\vspace{-4mm}
 \caption{On-the-fly decoding; removed edges are dashed: (a) Trellis after the reception of the source bit $b_1 = 0$, $(b)$ Trellis after the reception of the
 parity bit $b_2 = 1$, $(c)$ Trellis after left and right propagation. \vspace{-2mm}}
 \label{fig:decoding}
\end{figure}

\smallskip\noindent{\bf 6) Decoding stop}.
 The decoding is successful if $M_1 \left( i \right) \subseteq T_{b\x}$ for all $i \in \{0,\dots,K-1 \}$.  We then define
the inefficiency ratio $\mu$ as follows:
\begin{equation*}
\mu = \frac{r_{\mbox{\scriptsize stop}}}{K}
\end{equation*}
where $r_{\mbox{\scriptsize stop}} \geq K$ is the number of bits
received at the moment when the decoding stops.
 An illustration of the proposed algorithm is shown in Fig.
\ref{fig:decoding}. First, at the reception of an information bit
$b_1=0$, we remove the transitions in the corresponding step in the
trellis where $b_1=1$. Note that this step is done in interleaved
positions in both trellises at the reception of an information bit.
At this stage, no propagation in the trellis is possible as all the
states are still connected. Next we receive a parity bit $b_2=1$;
the remaining transitions corresponding to $b_2=0$ are removed.  At
that point, we notice that state $e_1$ and $e_2$ on the left are not
connected. This means that the transitions arriving from the left to
these states are not allowed anymore, thus they are removed.
Similarly, we remove the transitions leaving the states $e_1$ and
$e_3$ on the right.

In fact, the average decoding inefficiency $\mu_{av}$ of the code
relates to its erasure recovery capacity as follows: suppose that,
on average, the proposed decoding algorithm requires $K'$ ($K' \geq
K$) symbols to be able to recover the $K$ information symbols. we
can write the following:
\begin{equation*}
\mu_{av} = \frac{K'}{K} = \frac{\left(1- p_{th} \right) N}{K}=\frac{1-p_{th}}{R_c}
\end{equation*}
where the threshold probability $p_{th}$ corresponds to the average
fraction of erasures the decoder can recover. We can then write
$p_{th}$ as:
\begin{equation*}
p_{th} = 1 - \mu_{av} R_c
\end{equation*}
For instance, if a code with $R_c=1/3$ has $\mu_{av}=1.076$, it has
$p_{th}=0.641$. As a code with this coding rate is -theoretically-
capable of correcting a probability of erasure of $p=2/3$, the gap
to capacity is:
\begin{equation*}
\Delta_p = p - p_{th} \simeq 0.025
\end{equation*}
With codes such as LDPC or turbo codes, it is possible to achieve
near-capacity performance with iterative decoding, with $\mu_{av}
\simeq 1$. Ideally, an MDS code (that achieves capacity) has
$\mu_{av}=1$, {\em i.e.} it is capable of recovering the $K$
information symbols from any $K$ received symbols out of the $N$
codeword symbols.

Finally, it is important to note that the
algorithm proposed in this section is linear in the interleaver size
$K$.  In fact, an RSC code with $2^{ L-1}$
states and $2^k$ transitions leaving each states has $2^{
L-1} \times 2^k = 2^{k+L-1}$ transitions between two trellis
steps. This means that the turbo code has a total of approximately
$2 \times K \times 2^{k+L-1}$ transitions. Even if the decoding is
exponential in $k$ and $L$, it is linear in $K$.  As we can obtain
very powerful turbo codes with relatively small $k$ and $L$, we can
say that a turbo code with the proposed algorithm has linear time
encoding and decoding, and thus it is suited for applications were
low-complexity ``on-the-fly'' encoding/decoding are required (as
with the ``Raptor codes'' \cite{shokrollahi2006rc} for instance).
\section{Simulation Results}
\label{results}
 In this section, the performance of the proposed
algorithm with parallel turbo codes is shown. The coding rate of the
turbo code using half-rate constituent codes is $R_c=1/3$.  However,
we also consider turbo codes with $R_c=1/2$ and $R_c=2/3$ obtained
by puncturing the $R_c=1/3$ turbo code.  We use two types of
interleavers: 1) Pseudo-random (PR) interleavers (not optimized) and
2) Quasi-cyclic (QC) bi-di\-men\-sio\-nal interleavers
\cite{boutros_quasi-cyclic_2006} that are the best known
interleavers in the literature: in fact, it was shown in
\cite{breiling_logarithmic_2004} that the minimum distance $d_{min}$
of a turbo code is upper-bounded by a quantity that grows
logarithmically with the interleaver size $K$, and the QC
interleavers always achieve this bound.

The comparison is made with regular and irregular {\em staircase}
LDPC codes. An LDPC code is said to be {\em staircase} if the right
hand side of its parity check matrix consists of a double diagonal.
The advantage of a staircase LDPC code is that the encoding can be
performed in linear time using the parity check matrix, therefore
there is no need for the generator matrix, which generally is not
low density. A staircase LDPC code is said to be regular if the left
hand side of the parity check matrix is regular, {\em i.e.} the
number of $1$'s per column is constant. Otherwise it is said to be
irregular. In this section, we consider regular staircase LDPC codes
with four $1$'s per each left hand side column. Irregular staircase
LDPC codes are optimized for the BEC channel by density evolution.
\begin{figure}[!b]
\includegraphics[width=0.7\columnwidth,angle=-90]{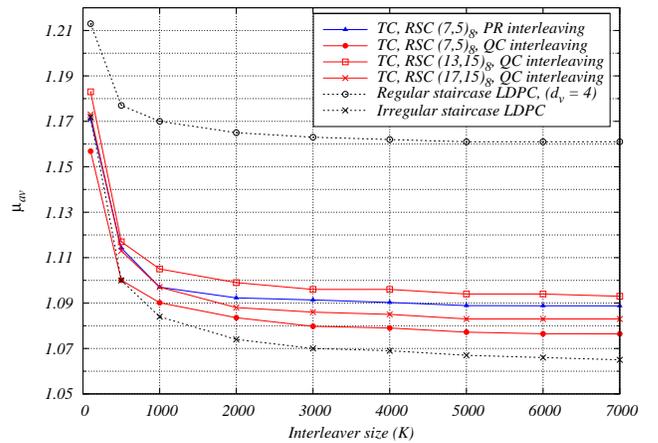}
\caption{Average inefficiency ($\mu_{av}$) with respect to
interleaver size $K$ over the BEC. Turbo code with half-rate RSC
constituents versus LDPC codes, $R_c=1/3$.} \label{Rc_1_3}
\end{figure}
In Fig. \ref{Rc_1_3}, we compare the performance of turbo codes and
LDPC codes for $R_c=1/3$. Turbo codes with RSC $(7,5)_8$ and PR
interleaving achieve an average inefficiency $\mu_{av}$ of about
$1.09$, which means they require $K'=1.09K$ received bits (or $9\%$
overhead) to be able to recover the $K$ information bits.  However,
using a QC interleaver, the overhead with the same turbo code is of
about $7.6\%$, which is very close to the irregular staircase LDPC
code, while the regular staircase LDPC is far above regular turbo
codes ($16\%$ overhead). In addition, it is important to note that
using turbo codes with RSC constituents with $L=4$ (trellis with
eight states), namely the RSC $(13,15)_8$ and the $(17,15)_8$ codes,
increases the decoding complexity without improving the performance.
\begin{figure}[t]
\includegraphics[width=0.7\columnwidth,angle=-90]{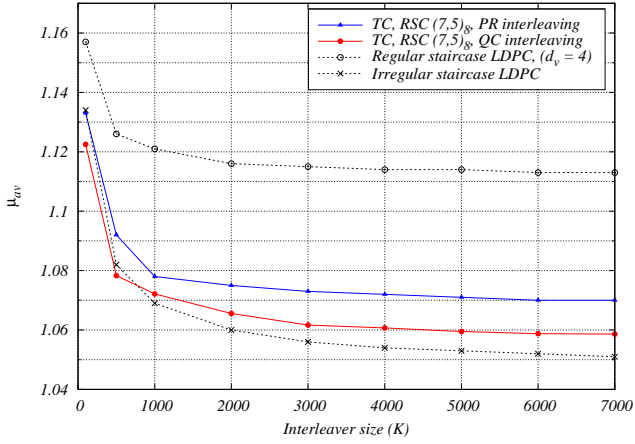}
\caption{Average inefficiency ($\mu_{av}$) with respect to
interleaver size $K$ over the BEC. Turbo code with half-rate
(7,5)$_{8}$ RSC constituents versus LDPC codes, punctured to
$R_c=1/2$.} \label{Rc_1_2}
\end{figure}
Punctured half-rate turbo codes are compared with half-rate LDPC
codes in Fig. \ref{Rc_1_2}.  Again, turbo codes with QC interleavers
largely outperform regular LDPC codes ($6\%$ to $11\%$ overhead),
and thus perform closer to irregular LDPC codes ($5\%$ overhead).
However, puncturing even more the turbo code to raise it to
$R_c=2/3$ widens the gap with irregular LDPC codes, placing the
performance curve with QC interleaving ($5.2\%$ overhead) at equal
distance from regular LDPC codes ($7.5\%$) and irregular LDPC codes
($3\%$).
\begin{figure}[t]
\includegraphics[width=0.7\columnwidth,angle=-90]{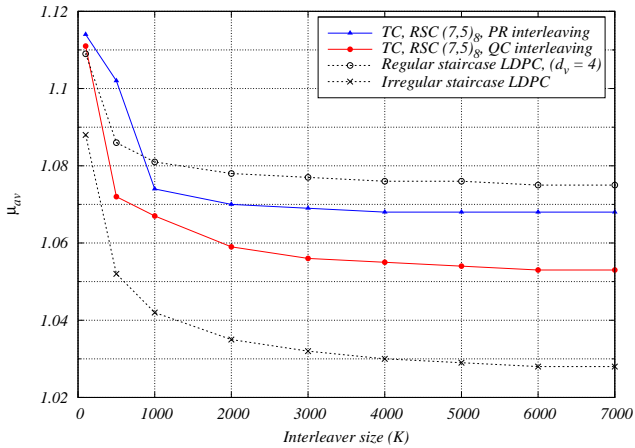}
\caption{Average inefficiency ($\mu_{av}$) with respect to
interleaver size $K$ over the BEC. Turbo code with half-rate
(7,5)$_{8}$ RSC constituents versus LDPC codes, punctured to
$R_c=2/3$.} \label{Rc_2_3}
\end{figure}

\section{Conclusion}
\label{conclusion} In this paper we proposed a novel decoding
algorithm for turbo codes over the BEC.  This algorithm,
characterized by ``on-the-fly'' propagation in the trellises and
hard information exchange between the two codes, is appropriate for
UL-FEC. Performance results with very small overhead were shown for
different interleaver sizes and coding rates. Although the turbo
codes presented in this paper were not optimized for the BEC, the
results are very promising.  Further improvements can be done by
optimizing turbo codes for this channel.

\bibliographystyle{IEEEbib}
\footnotesize
\bibliography{strings,zotero}

\begin{thebibliography}{10}

\bibitem{elias1955cnc}
P.~Elias,
\newblock ``{Coding for noisy channels},''
\newblock {\em IRE Conv. Rec}, vol. 4, no. S 37, pp. 47, 1955.

\bibitem{di2002fla}
C.~Di, D.~Proietti, E~Telatar, T~Richardson, and R~Urbanke,
\newblock ``{Finite-length analysis of low-density parity-check codes on
  thebinary erasure channel},''
\newblock {\em IEEE Trans. Inf. Theory}, vol. 48, no. 6, pp. 1570--1579, 2002.

\bibitem{richardson_finite-length_2002}
T.~Richardson, A.~Shokrollahi, and R.~Urbanke,
\newblock ``Finite-length analysis of various low-density parity-check
  ensembles for the binary erasure channel,''
\newblock {\em IEEE Int. Symp. Inf. Theory}, 2002.

\bibitem{richardson_design_2001}
T.~Richardson, A.~Shokrollahi, and R.~Urbanke,
\newblock ``Design of capacity-approaching irregular low-density parity-check
  codes,''
\newblock {\em IEEE Trans. Inf. Theory}, vol. 47, pp. 619--637, 2001.

\bibitem{luby2001eec}
M.G. Luby, M.~Mitzenmacher, M.A. Shokrollahi, and D.A. Spielman,
\newblock ``{Efficient erasure correcting codes},''
\newblock {\em IEEE Trans. Inf. Theory}, vol. 47, no. 2, pp. 569--584, 2001.

\bibitem{pfister2005cae}
HD~Pfister, I.~Sason, and R.~Urbanke,
\newblock ``{Capacity-achieving ensembles for the binary erasure channel with
  bounded complexity},''
\newblock {\em IEEE Trans. Inf. Theory}, vol. 51, no. 7, pp. 2352--2379, 2005.

\bibitem{luby2002lc}
M.~Luby,
\newblock ``{LT codes},''
\newblock {\em Proc. ACM Symp. Found. Comp. Sci.}, pp. 271--280, 2002.

\bibitem{shokrollahi2006rc}
A.~Shokrollahi,
\newblock ``{Raptor codes},''
\newblock {\em IEEE/ACM Trans. Networking (TON)}, vol. 14, pp. 2551--2567,
  2006.

\bibitem{berrou_near_1996}
C.~Berrou and A.~Glavieux,
\newblock ``Near optimum error correcting coding and decoding: turbo-codes,''
\newblock {\em IEEE Trans. Comm.}, vol. 44, pp. 1261--1271, 1996.

\bibitem{tepe_turbo_1998}
K.E. Tepe and J.B. Anderson,
\newblock ``Turbo codes for binary symmetric and binary erasure channels,''
\newblock {\em IEEE Int. Symp. Inf. Theory}, 1998.

\bibitem{kurkoski_exact_2003}
B.M. Kurkoski, P.H. Siegel, and J.K. Wolf,
\newblock ``{Exact probability of erasure and a decoding algorithm for
  convolutional codes on the binary erasure channel},''
\newblock {\em IEEE GLOBECOM}, 2003.

\bibitem{kurkoski_analysis_2004}
B.M. Kurkoski, P.H. Siegel, and J.K. Wolf,
\newblock ``Analysis of convolutional codes on the erasure channel,''
\newblock {\em IEEE Int. Symp. Inf. Theory}, 2004.

\bibitem{rosnes_turbo_2007}
E.~Rosnes and O.~Ytrehus,
\newblock ``Turbo decoding on the binary erasure channel: Finite-length
  analysis and turbo stopping sets,''
\newblock {\em IEEE Trans. Inf. Theory}, vol. 53, pp. 4059--4075, 2007.

\bibitem{jeong_w_lee_performance_2007}
Jeong~W. Lee, R.~Urbanke, and R.E. Blahut,
\newblock ``On the performance of turbo codes over the binary erasure
  channel,''
\newblock {\em IEEE Comm. Lett.}, vol. 11, pp. 67--69, 2007.

\bibitem{bcjr}
L.~Bahl, J.~Cocke, F.~Jelinek, and J.~Raviv,
\newblock ``{Optimal decoding of linear codes for minimizing symbol error
  rate},''
\newblock {\em IEEE Trans. Inf. Theory}, vol. 20, no. 2, pp. 284--287, March
  1974.

\bibitem{viterbi_error_1967}
A.~Viterbi,
\newblock ``Error bounds for convolutional codes and an asymptotically optimum
  decoding algorithm,''
\newblock {\em IEEE Trans. Inf. Theory}, vol. 13, pp. 260--269, 1967.

\bibitem{boutros_quasi-cyclic_2006}
J.J. Boutros and G.~Zemor,
\newblock ``On quasi-cyclic interleavers for parallel turbo codes,''
\newblock {\em IEEE Trans. Inf. Theory}, vol. 52, pp. 1732--1739, 2006.

\bibitem{breiling_logarithmic_2004}
M.~Breiling,
\newblock ``A logarithmic upper bound on the minimum distance of turbo codes,''
\newblock {\em IEEE Trans. Inf. Theory}, vol. 50, pp. 1692--1710, 2004.

\end{thebibliography}

\end{document}